\documentclass[10pt, conference, compsocconf]{IEEEtran}
\IEEEoverridecommandlockouts
\usepackage{paralist}
\usepackage{algorithm}
\usepackage{algorithmic}

\usepackage{amsmath}
\usepackage{multirow}

\usepackage{subfigure}
\usepackage{graphicx}
\usepackage{epsfig}
\usepackage{epstopdf}

\usepackage{amssymb}

\DeclareMathOperator*{\argmin}{argmin}

\begin{document}
\newtheorem{definition}{Definition}


\title{\Large Collaborative Inference of Coexisting Information Diffusions}

\newcommand{\superscript}[1]{\ensuremath{^{\textrm{#1}}}}
\def\sharedaffiliation{\end{tabular}\newline\begin{tabular}{c}}

\def \scu{\superscript{*}}
\def \uic{\superscript{\dag}}

\DeclareRobustCommand*{\IEEEauthorrefmark}[1]{%
  \raisebox{0pt}[0pt][0pt]{\textsuperscript{\footnotesize\ensuremath{#1}}}}

\author{\IEEEauthorblockN{Yanchao Sun\IEEEauthorrefmark{1},
Cong Qian\IEEEauthorrefmark{1},
Ning Yang\IEEEauthorrefmark{1}\IEEEauthorrefmark{*} \thanks{\IEEEauthorrefmark{*} Ning Yang is the corresponding author.}, 
Philip S. Yu\IEEEauthorrefmark{2}}
\IEEEauthorblockA{\IEEEauthorrefmark{1}School of Computer Science, Sichuan University,
Chengdu, China\\ Email: \{sunyc, congqian\}@st.scu.edu.cn, yangning@scu.edu.cn}
\IEEEauthorblockA{\IEEEauthorrefmark{2}Department of Computer Science, University of Illinois at Chicago, Chicago, USA\\
Email: psyu@uic.edu}}

\maketitle
\begin{abstract}
Recently, \textit{diffusion history inference} has become an emerging research topic due to its great benefits for various applications, whose purpose is to reconstruct the missing histories of information diffusion traces according to incomplete observations. The existing methods, however, often focus only on single information diffusion trace, while in a real-world social network, there often coexist multiple information diffusions over the same network. In this paper, we propose a novel approach called Collaborative Inference Model (CIM) for the problem of the inference of coexisting information diffusions. By exploiting the synergism between the coexisting information diffusions, CIM holistically models multiple information diffusions as a sparse 4th-order tensor called Coexisting Diffusions Tensor (CDT) without any prior assumption of diffusion models, and collaboratively infers the histories of the coexisting information diffusions via a low-rank approximation of CDT with a fusion of heterogeneous constraints generated from additional data sources. To improve the efficiency, we further propose an optimal algorithm called Time Window based Parallel Decomposition Algorithm (TWPDA), which can speed up the inference without compromise on the accuracy by utilizing the temporal locality of information diffusions. The extensive experiments conducted on real world datasets and synthetic datasets verify the effectiveness and efficiency of CIM and TWPDA.
\end{abstract}



\IEEEpeerreviewmaketitle

\section{Introduction}

Recently, \textit{diffusion history inference} has become an emerging research topic of which the purpose is to reconstruct the missing histories of information diffusion traces according to incomplete observations \cite{r47}. Although information diffusions are ubiquitous in social networks, it is hard to know the whole history of a diffusion, due to privacy or physical limitations. However, learning the whole history of a diffusion plays an important role in many applications. For example, we likely can not identify a rumor propagation over a network until it is noticed that a significant number of users have transmitted it. In this case, it is essential to learn the diffusion history of the rumor for stopping its future spreading.

Some approaches \cite{r47,r48,r25} have been proposed for the problem of diffusion history inference. The existing methods, however, often focus only on single information diffusion trace, while in a real-world social network, there often coexist multiple diffusions of different information, and a node often participates in more than one of the information diffusions at the same time period. Fig. \ref{Fig_Motivation} shows an example of coexisting information diffusions in twitter, where a diffusion of meme 1 (represented by blue solid arrows) and a diffusion of meme 2 (represented by red dashed arrows) coexist in the same network during 4 time points (from $t_1$ to $t_4$). 


In real world, available observations of a single information diffusion are usually incomplete and sparse \cite{r05,r53}. For example, in Fig. \ref{Fig_Motivation}, at each time point, only a very few infections of memes 1 and 2 are observed. If the infection of meme 2 from $v_2$ to $v_7$ at $t_1$ and the infection of meme 1 from $v_8$ to $v_5$ at $t_2$ are not observed, can we recover them at the same time? In this paper, we investigate the problem of \textit{using sparse observations to infer the detailed histories of multiple information diffusions coexisting in a given network}, which is not easy due to the following challenges:


\begin{compactenum}[(1)]

\item \textbf{Data Sparsity} 
Although massive data for information diffusions have been collected in the big data era, available observations of a single information diffusion are still likely sparse and insufficient, which makes it extremely hard to infer the coexisting information diffusions separately. 

\item \textbf{Lack of Priori Knowledge} The existing works often assume that information diffusion follows a parametric model such as Independent Cascade (IC) model and Susceptible-Infected (SI) model. In real world, however, information diffusion processes are so complicated that we seldom exactly know how information diffuses \cite{r25}. 

\item \textbf{Algorithmic Parallelizability} In the big data era, a network often consists of billions of nodes and edges. Parallelizable algorithms are indispensable to the inference of coexisting information diffusions on a huge network. 
\end{compactenum}

\begin{figure}[tb]
\centering
    \epsfig{file=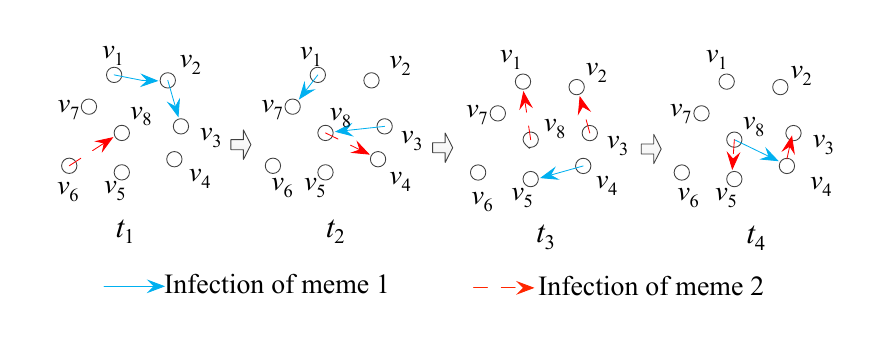}
\caption{Illustration of Coexisting Information Diffusions with Sparse Observations.}
\label{Fig_Motivation}
\end{figure}

In this paper, we propose a novel model, called Collaborative Inference Model (CIM) for the problem of the inference of coexisting information diffusions. By exploiting the synergism between the coexisting information diffusions, CIM holistically models the coexisting information diffusions as a 4th-order tensor, called Coexisting Diffusions Tensor (CDT), without priori assumption of diffusion models, and fulfills the collaborative inference via a low-rank approximation of sparse CDT. To improve the inference accuracy, four heterogeneous constraints, which can be generated from additional data sources, are fused with the decomposition. By plugging in the heterogeneous constraints into the estimation of CDT, the knowledge obtained from additional data sources are transferred to the target social network in which coexisting information diffusions will be inferred. 

To address the challenge of algorithmic scalability, we further propose an optimal decomposition algorithm, called Time Window based Parallel Decomposition Algorithm (TWPDA), as the core component of CIM to fulfill the approximation of CDT. By exploiting the temporal locality of information diffusions, TWPDA can speed up the decomposition in a parallel way without loss of approximation accuracy . 

Our contributions can be summarized as follows:

\begin{compactenum}[(1)]


\item We propose a novel model, called Collaborative Inference Model (CIM) for the inference of coexisting information diffusions from sparse observations. CIM models coexisting information diffusions as a CDT without priori assumption of diffusion models, and can collaboratively infer the missing observations via a low-rank approximation of sparse CDT with a fusion of heterogeneous constraints generated from additional data sources.

\item  We propose an optimal algorithm, called Time Window based Parallel Decomposition Algorithm (TWPDA), to speed up the decomposition of a sparse CDT without loss of accuracy by utilizing the temporal locality of information diffusions.

\item Extensive experiments are conducted on real datasets to verify the effectiveness and efficiency of the proposed approach.

\end{compactenum}

The remainder of this paper is organized as follows. The preliminaries and basic definitions are introduced in Section 2. The details of CIM are described in Section 3. The details of TWPDA are presented in Section 4. We analyze the experimental results in Section 5. At last, we briefly review the related work in Section 6 and conclude in Section 7.

\section{Preliminaries}

\subsection{Basic Definitions}

Let $C = \{ c_1, \cdots, c_M\}$ be the set of $M$ memes whose diffusions coexist over a network $G$ consisting of $N$ nodes $\{v_1, \cdots, v_N\}$ during $Q$ time points $\langle t_1, t_2, \cdots, t_Q \rangle$, where the term \textit{meme} represents anything that can propagate over a social network, for example, it can be a label, a key word, or a news, etc.

\begin{definition}
An \textit{infection} $e$ is a 4-tuple $(e.s, $ $e.d, e.c, e.t)$, which represents an infection of meme $c \in C$ from source node $s$ to destination node $d$ ($s, d \in V$) at time point $t$.
\end{definition}

In our case, term \textit{infection} is a generalized concept which represents the smallest blocks constituting a diffusion trace. For example,  when one user posts a tweet of a meme, we can claim the users following that user are infected by that meme from that user if they react to that tweet (comment on or retweet that tweet). Note that an infection can happen more than once. For example, one may post tweets of the same meme many times in the same week.
\begin{definition}
  \textbf{Information Diffusion}: An information diffusion of meme $c_m$ ($m = 1,2, \cdots, M$) is a set $E^{(m)}$ consisting of infections whose memes are $c_m$, i.e., $E^{(m)} = \{e|e.c = c_m\}$.
\end{definition}

Note that an information diffusion $E^{(m)}$ can be viewed as a sequence of its temporal snapshots $E^{(m)}_q$ ($q = 1, \cdots, Q$), i.e., $E^{(m)} = \bigcup_{q=1}^{Q}E^{(m)}_q$, where $E^{(m)}_q$ is the snapshot at time point $t_q$, i.e., $E^{(m)}_q = \{e| $ $e \in E^{(m)} \text{ and } e.t = t_q \}$. 

\subsection{Problem Statement}

Given a collection of sparse observations of coexisting information diffusions of $M$ memes, $\{{O}^{(1)}, {O}^{(2)}, \cdots, {O}^{(M)} \}$, where ${O}^{(m)}$ is the observation of $E^{(m)}$, $m = 1, 2, \cdots, M$, i.e., ${O}^{(m)} \subset E^{(m)}$, we want to infer the complete coexisting information diffusions $\{\widehat{E}^{(1)}, \widehat{E}^{(2)}, \cdots, $ $\widehat{E}^{(M)} \}$.

\section{Collaborative Inference Model}

\subsection {Coexisting Diffusions Tensor}

We build the Coexisting Diffusions Tensor (CDT) with the sparse observations of $M$ coexisting information diffusions, $\{{O}^{(1)}, {O}^{(2)}, \cdots, {O}^{(M)} \}$. A CDT $\boldsymbol{\mathcal{A}}\in \mathbb{R}^{N\times N\times M \times Q}$ consists of 4 modes which respectively represent $N$ source nodes, $N$ destination nodes, $M$ memes that concurrently diffuse, and $Q$ time points. A cell of the CDT, $\boldsymbol{\mathcal{A}}_{ijmq}$, stores the frequency of the occurrences of an infection $e$ with $e.s = v_i$, $e.d = v_j, e.c = c_m, e.t = t_q$, i.e., the infection of meme $c_m$ from source node $v_i$ to destination node $v_j$ at time point $t_q$, where $i$, $j  = 1, 2, \cdots, N$, $m = 1, 2, \cdots, M$, and $q = 1, 2, \cdots, Q$. A time slice at $t_q$ is a 3-dimensional tensor $\boldsymbol{\mathcal{A}}_{***q} \in \mathbb{R}^{N\times N\times M}$, which can be assembled with the sequence of snapshots of $M$ coexisting diffusions at $t_q$, i.e., $\{{O}_q^{(1)}, {O}_q^{(2)}, \cdots, {O}_q^{(M)} \}$. Note that in a CDT, the set of source nodes is the same as the set of destination nodes, since one can be a source and a destination as well. 

\subsection{Sparse CDT Approximation}
As CDT codes all the information about the coexisting diffusions, the problem of the inference of coexisting information diffusions can be reduced to the approximation of the sparse CDT $\boldsymbol{\mathcal{A}}$. We construct a dense tensor $\boldsymbol{\widehat{\mathcal{A}}}$ as the estimate of $\boldsymbol{\mathcal{A}}$ by a Tucker decomposition \cite{r06},
\begin{equation}
\boldsymbol{\mathcal{A}} \approx \boldsymbol{\widehat{\mathcal{A}}} = \boldsymbol{\mathcal{G}} \times_1 \boldsymbol{S}  \times_2 \boldsymbol{D}  \times_3 \boldsymbol{C} \times_4 \boldsymbol{T},
\label{Eq_CDT_Decomposition}
\end{equation}
i.e., a core tensor $\boldsymbol{\mathcal{G}} \in \mathbb{R}^{R \times R \times R \times R}$ multiplied by 4 latent factor matrices, $\boldsymbol{S} \in \mathbb{R}^{N \times R}$, $\boldsymbol{D} \in \mathbb{R}^{N \times R}$, $\boldsymbol{C} \in \mathbb{R}^{M \times R}$, and $\boldsymbol{T} \in \mathbb{R}^{Q \times R}$, along its 4 modes respectively, where $R$ is the target rank, and the symbol $\times_{i}$ ($1 \le i \le 4$) stands for the tensor multiplication along the $i$th mode. One can note that the decomposition expressed by Equation (\ref{Eq_CDT_Decomposition}) is meaningful. Essentially, $\boldsymbol{S}$, $\boldsymbol{D}$, $\boldsymbol{C}$, and $\boldsymbol{T}$ can be regarded as the latent feature factors about source nodes, destination nodes, memes, and time, respectively.

To achieve a higher approximation accuracy, $\boldsymbol{\widehat{A}}$ is produced by solving the following optimization problem:
\begin{equation}
\argmin_{\boldsymbol{\mathcal{G}}, \boldsymbol{S}, \boldsymbol{D}, \boldsymbol{C}, \boldsymbol{T}} \mathcal{L}(\boldsymbol{\mathcal{G}}, \boldsymbol{S}, \boldsymbol{D}, \boldsymbol{C}, \boldsymbol{T}).
\label{Eq_CIM}
\end{equation}
The objective function $\mathcal{L}(\boldsymbol{\mathcal{G}}, \boldsymbol{S}, \boldsymbol{D}, \boldsymbol{C}, \boldsymbol{T})$ is defined as
\begin{equation}
\begin{split}
\mathcal{L}(\boldsymbol{\mathcal{G}}, \boldsymbol{S}, \boldsymbol{D}, \boldsymbol{C}, \boldsymbol{T}) & = \frac{1}{2} \epsilon(\boldsymbol{\widehat{\mathcal{A}}}) + \frac{\lambda_1}{2}\gamma(\boldsymbol{\mathcal{G}}, \boldsymbol{S}, \boldsymbol{D}, \boldsymbol{C}, \boldsymbol{T}) \\
& +\frac{\lambda_2}{2} \phi(\boldsymbol{S}, \boldsymbol{D}) + \frac{\lambda_3}{2} \psi(\boldsymbol{D}, \boldsymbol{C}) \\
& +\frac{\lambda_4}{2} \xi(\boldsymbol{C}) + \frac{\lambda_5}{2} \tau(\boldsymbol{T}),
\end{split}
\label{Eq_Objective}
\end{equation}
where $\lambda_1 \thicksim \lambda_5$ are the nonnegative parameters used to control the respective contributions of the terms. 

Equations (\ref{Eq_CDT_Decomposition}), (\ref{Eq_CIM}), and (\ref{Eq_Objective}) together define our CIM model. In Equation (\ref{Eq_Objective}), the first term $\epsilon(\boldsymbol{\widehat{\mathcal{A}}})$ is the reconstruction error of the observable cells, which is defined as
\begin{equation}
\epsilon(\boldsymbol{\widehat{\mathcal{A}}}) = \Arrowvert  \boldsymbol{\widehat{\mathcal{A}}}_\Omega - \boldsymbol{\mathcal{A}}_\Omega \Arrowvert_F^2,
\label{Eq_Error}
\end{equation}
where $\Arrowvert \cdot \Arrowvert_F$ stands for Frobenius norm of tensor, and $\Omega$ represents the set of the indices of the observable tensor cells. The second term $\gamma(\boldsymbol{\mathcal{G}}, \boldsymbol{S}, \boldsymbol{D}, \boldsymbol{C}, \boldsymbol{T})$ is the regularization constraints for avoiding overfitting, which is defined as
\begin{equation}
\gamma(\boldsymbol{\mathcal{G}}, \boldsymbol{S}, \boldsymbol{D}, \boldsymbol{C}, \boldsymbol{T}) =\Arrowvert \boldsymbol{\mathcal{G}}\Arrowvert_2^2 + \Arrowvert  \boldsymbol{S}\Arrowvert_2^2  + \Arrowvert  \boldsymbol{D} \Arrowvert_2^2 + \Arrowvert  \boldsymbol{C}\Arrowvert_2^2  + \Arrowvert  \boldsymbol{T} \Arrowvert_2^2.
\label{Eq_Regularization}
\end{equation}
where $\Arrowvert \cdot \Arrowvert_2$ stands for 2-norm of matrix.

To improve the approximation accuracy, we decompose a sparse CDT with a fusion of four additional constraints, i.e., the Source-Destination Affinity (DSA), the Node-Meme Affinity (NMA), the Meme Correlation (MC), and the Temporal Smoothness (TS). These constraints are respectively represented by the last four terms of Equation (\ref{Eq_Objective}), i.e., $\phi(\boldsymbol{S}, \boldsymbol{D})$, $\psi(\boldsymbol{D}, \boldsymbol{C})$, $\xi(\boldsymbol{C})$, and $\tau(\boldsymbol{T})$, which can be generated from additional data sources, as described in the following subsection.

\subsection{Heterogeneous Constraints}
\subsubsection {Source-Destination Affinity Constraint}
In social networks, whether an infection happens partly depends on the affinity between the source node and destination node, and the affinity strength is not reciprocal \cite{r40,r07,r08}. For example, Alice may react to Bob more frequently than Bob reacts to her. Based on this idea, we build a Source-Destination Affinity (SDA) matrix $\boldsymbol{X} \in \mathbb{R}^{N \times N}$ from an additional data source. $\boldsymbol{X}$ is asymmetric, and an element $\boldsymbol{X}_{ij}$ indicates how likely node $v_i$ will react to its neighbor node $v_j$, $1 \le i, j \le N$. For example, in twitter, larger the $\boldsymbol{X}_{ij}$, more likely $v_i$ will retweet the tweets of $v_j$. Actually, to build $\boldsymbol{X}$, we need to rank the neighnor nodes of $v_i$ in terms of their importance to $v_i$. For this purpose, we define $\boldsymbol{X}_{ij}$ as 
\begin{equation}
\boldsymbol{X}_{ij} = \frac{\arrowvert \{v_p\mid f_{ip}<f_{ij} \} \arrowvert + 0.5 \arrowvert \{v_p\mid f_{ip}=f_{ij} \} \arrowvert} {\arrowvert F_i \arrowvert},
\label{Eq_Affinity_Matrix}
\end{equation}
where $F_i$ is the set of the neighbors of $v_i$, $v_p \in F_i$, and $f_{ij}$ is the number of reactions of $v_i$ to $v_j$. It is easy to show that $0 < \boldsymbol{X}_{ij} \le 1$. Note that the first term of the numerator is the number of nodes who receive reactions from $v_i$ less than $v_j$ does, while the second term is half of the number of nodes who receive reactions from $v_i$ as many as $v_j$ does. So $\boldsymbol{X}_{ij}$ can be regarded as the ratio of the nodes to which $v_i$ will react less likely than or as likely as to $v_j$, and higher $\boldsymbol{X}_{ij}$ reasonably indicates $v_j$ is of more importance than $v_i$. If $\boldsymbol{X}_{ij} = 1$, no neighbor node receives more reactions from $v_i$ than $v_j$ does, which implies that $v_j$ is the one $v_i$ cares most.

To fuse $\boldsymbol{X}$ with the decomposition, we factorize it as $\boldsymbol{X} = \boldsymbol{S} \times \boldsymbol{D}^T$, where $\boldsymbol{S}$ and $\boldsymbol{D}$ are exactly the latent feature matrices of source nodes and destination nodes of infections, and shared with the CDT decomposition shown in Equation \ref{Eq_CDT_Decomposition}. Thus the SDA constraint, $\phi(\boldsymbol{S}, \boldsymbol{D})$, is defined as
\begin{equation}
\phi(\boldsymbol{S}, \boldsymbol{D}) = \Arrowvert \boldsymbol{X} - \boldsymbol{S} \boldsymbol{D}^T\Arrowvert_2^2.
\label{Eq_DSA}
\end{equation}
The insight here is that the knowledge of coexisting information diffusions, which is represented by the CDT $\boldsymbol{\widehat{\mathcal{A}}}$, can fuse, through $\boldsymbol{S}$ and $\boldsymbol{D}$, with the social affinity features of the nodes, which is represented by $\boldsymbol{X}$.

\subsubsection{Node-Meme Affinity Constraint}
Intuitively, one likely cares more about some specific memes than other memes. For example, if we know that one pays closer attention to basketball games than to TV series, we can infer that the she/he has more potential for knowing a latest basketball game than knowing a new TV series. Thus the stronger affinity a node has with a meme, the more likely she/he is infected by it. Based on this idea, we build a Node-Meme Affinity (NMA) matrix $\boldsymbol{Y} \in \mathbb{R}^{N \times M}$ from an additional data source. An element $\boldsymbol{Y}_{ij}$ ($1 \le i \le N$, $1 \le j \le M$) is defined as the proportion of the infections with meme $c_j$ in the infections with destination $v_i$, i.e.,
\begin{equation}
 \boldsymbol{Y}_{ij} = \frac{\arrowvert \{ e\mid e.d = v_i \land e.c = c_j\}\arrowvert}{\arrowvert \{ e\mid e.d = v_i\}\arrowvert},
\label{Eq_NMA_Matrix}
\end{equation}
where $e$ is an infection. Essentially, as the NMA matrix $\boldsymbol{Y}$ carries the information of the destination nodes and memes of infections, we can factorize it as $\boldsymbol{Y} = \boldsymbol{D} \times \boldsymbol{C}^T$, where $\boldsymbol{D}$ and $\boldsymbol{C}$ are the latent feature matrices of destination nodes and memes, respectively. Note that $\boldsymbol{Y}$ shares $\boldsymbol{D}$ and $\boldsymbol{C}$ with Equation (\ref{Eq_CDT_Decomposition}), hence to fuse $\boldsymbol{Y}$ with the decomposition, the NMA constant is defined as 
\begin{equation}
\psi(\boldsymbol{D}, \boldsymbol{C}) = \Arrowvert \boldsymbol{Y} - \boldsymbol{D} \boldsymbol{C}^T\Arrowvert_2^2.
\label{Eq_NMA}
\end{equation}

\subsubsection{Meme-Correlation Constraint}
We often observe that related memes likely have similar diffusion pattern, i.e., one who is infected by a meme is also likely infected by a related meme. For example, one who is interested in meme 'basketball' may also pay more attention to meme 'football' than meme about Korean TV series, as 'basketball' and 'football' both fall into the category of sport. Based on this observation, to capture the correlation between memes, we build a Meme Correlation (MC) matrix $\boldsymbol{Z} \in \mathbb{R}^{M \times M}$ from additional data sources. An element $\boldsymbol{Z}_{ij}$ is defined as the proportion of co-occurrence of meme $c_i$ and $c_j$, i.e., 
\begin{equation}
\boldsymbol{Z}_{ij} = \frac{\arrowvert N_i \cap N_j \arrowvert} {\arrowvert N_i \arrowvert + \arrowvert N_j \arrowvert - \arrowvert N_i \cap N_j \arrowvert},
\label{Eq_MC_matrix}
\end{equation}
where $N_i = \{ v | v \in V \land e.d = v \land e.c = c_i\}$ is the set of nodes infected by meme $c_i$, and $N_j = \{ v | v \in V \land e.d = v \land e.c = c_j\}$ is the set of nodes infected by meme $c_j$. Reasonably, if $\boldsymbol{Z}_{ij}$ is large, the distance between meme latent feature vectors $\boldsymbol{C}_{i*}$ (the $i$-th row vector of $\boldsymbol{C}$) and $\boldsymbol{C}_{j*}$ (the $j$-th row vector of $\boldsymbol{C}$) should be small, which leads to the following MC constraint:
\begin{equation}
\xi(\boldsymbol{C}) = tr(\boldsymbol{C}^T (\boldsymbol{K}-\boldsymbol{Z}) \boldsymbol{C}),
\label{Eq_MC}
\end{equation}
where $\boldsymbol{K} \in \mathbb{R}^{M \times M}$ is a diagonal matrix with diagonal elements $\boldsymbol{K}_{ii} = \sum_{j}{\boldsymbol{Z}_{ij}}$. To see why $\xi(\boldsymbol{C})$ works, note that we can get $\xi(\boldsymbol{C}) = \frac{1}{2} \Arrowvert {\boldsymbol{C}_{i*}}^T - {\boldsymbol{C}_{j*}}^T\Arrowvert_2 \boldsymbol{Z}_{ij}$ after a simple derivation. Therefore, the first factor $\Arrowvert {\boldsymbol{C}_{i*}}^T - {\boldsymbol{C}_{j*}}^T\Arrowvert_2$ will be small if the second factor $\boldsymbol{Z}_{ij}$ is large, so that $\xi(\boldsymbol{C}) $ can reach minimum as required by the optimizing objective defined by Equations (\ref{Eq_CIM}) and (\ref{Eq_Objective}).

\subsubsection{Temporal Smoothness Constraint}
Like heat diffusion, information diffusions on networks always progress gradually without drastic change of the states of nodes \cite{r09,r10,r11,r12,r13}, which exhibits a property of Temporal Smoothness (TS) of information diffusion. Inspired by this observation, we impose on the CDT decomposition the TS constraint $\tau(\boldsymbol{T})$ which is defined as:
\begin{equation}
\tau(\boldsymbol{T}) =  \Arrowvert \boldsymbol{T} - \boldsymbol{U} \boldsymbol{T}\Arrowvert_2^2,
\label{Eq_TS}
\end{equation}
where $\boldsymbol{U} \in \mathbb{R}^{Q \times Q}$ is the temporal smoothing matrix with 0 elements except for $\boldsymbol{U}_{i,i+1}$ = 1, $1 \le i \le Q-1$. By minimizing $\tau(\boldsymbol{T})$, it is guaranteed that the temporal latent features at successive two time points are similar.

\section{Decomposition Algorithms}

\subsection{Native Decomposition Algorithm}

As there is no closed-form solution to Equation (\ref{Eq_CIM}), we first propose a Native Decomposition Algorithm (NDA) based on gradient descent to find a local optimum solution. The outline of NDA is given in Algorithm \ref{Alg_NDA}, where nonzero cells of the sparse CDT are updated along the gradient direction until the objective function converges (Lines 6 to 16). NDA performs the decomposition on the whole tensor (that is why it is called 'native'), which makes it unable to scale up to a big CDT with millions cells. 

The essential part of NDA is the loop from Line 6 to Line 16, where we update every factor of decomposition according to the value of each nonzero cell of CDT. The updating computation takes a constant time, so the computational complexity for each iteration of NDA is roughly $O(n)$, where $n$ is the number of nonzero elements of CDT.

\renewcommand{\algorithmicrequire}{\textbf{Input:}}
\renewcommand{\algorithmicensure}{\textbf{Output:}}

\begin{algorithm}[t]
  \caption{ \emph{NDA}$(\boldsymbol{\mathcal{A}},\boldsymbol{X},\boldsymbol{Y},\boldsymbol{Z},\boldsymbol{U},R,\epsilon)$ }
  \label{Alg_NDA}
  \begin{algorithmic}[1]
    \REQUIRE ~~ 
       $\boldsymbol{\mathcal{A}}$: the sparse CDT; 
       $\boldsymbol{X}$: the SDA matrix; 
       $\boldsymbol{Y}$: the NMA matrix; 
       $\boldsymbol{Z}$: the MC matrix; 
       $R$: the target rank; 
       $\epsilon$: the threshold of error;
    \ENSURE ~~ 
       $\boldsymbol{\widehat{\mathcal{A}}}$: the approximated CDT;
    \STATE Randomly initialize matrices $\boldsymbol{S}$, $ \boldsymbol{D}$, $\boldsymbol{C}$, $\boldsymbol{T}$, and core tensor $\boldsymbol{\mathcal{G}}$ ;
    \STATE Set $\eta$ as step size;
    \STATE Set $\boldsymbol{K}_{ii}=\sum_{i=1}^M \boldsymbol{Z}_{ij}$ for $1 \le i \le M$;
    \STATE $\boldsymbol{L}_Z=\boldsymbol{K}-\boldsymbol{Z}$;
    \STATE $loss_a = \mathcal{L}(\boldsymbol{\mathcal{G}}, \boldsymbol{S}, \boldsymbol{D}, \boldsymbol{C}, \boldsymbol{T})$;
    \REPEAT
        \FORALL {$\boldsymbol{\mathcal{A}}_{ijkl}\neq 0$}
           \STATE $\boldsymbol{\mathcal{B}}_{ijkl} = \boldsymbol{\mathcal{G}}\times_1 \boldsymbol{S}_{i*} \times_2 \boldsymbol{D}_{j*} \times_3 \boldsymbol{C}_{k*} \times_4 \boldsymbol{T}_{l*}$;
           
           \STATE $\boldsymbol{S}_{i*} = \boldsymbol{S}_{i*} - \eta\lambda_5 \boldsymbol{S}_{i*} - \eta\lambda_1(\boldsymbol{S}_{i*}\times \boldsymbol{D}^T - \boldsymbol{X}_{i*})\times \boldsymbol{D}$\\
           \qquad \qquad $ - \eta (\boldsymbol{\mathcal{B}}_{ijkl} - \boldsymbol{\mathcal{A}}_{ijkl})\times \boldsymbol{\mathcal{G}} \times_2 \boldsymbol{D} \times_3 \boldsymbol{C} \times_4 \boldsymbol{T}$;
           
           \STATE $\boldsymbol{D}_{j*} = \boldsymbol{D}_{j*} - \eta\lambda_5 \boldsymbol{D}_{j*} - \eta\lambda_1(S\times \boldsymbol{D}_{j*}^T - \boldsymbol{X}_{*j})^T\times \boldsymbol{S} $\\
           \qquad \qquad $ - \eta\lambda_2 ( \boldsymbol{D}_{j*}\times \boldsymbol{C}^T - \boldsymbol{Y}_{j*})\times \boldsymbol{C} $\\
           \qquad \qquad $ - \eta (\boldsymbol{\mathcal{B}}_{ijkl} - \boldsymbol{\mathcal{A}}_{ijkl})\times \boldsymbol{\mathcal{G}} \times_1 \boldsymbol{S} \times_3 \boldsymbol{C} \times_4 {T}$;
           
           \STATE $\boldsymbol{C}_{k*} = \boldsymbol{C}_{k*} - \eta\lambda_5 \boldsymbol{C}_{k*} - \eta\lambda_2 (\boldsymbol{D} \times \boldsymbol{C}_{k*}^T - \boldsymbol{Y}_{*k})^T\times \boldsymbol{D} $\\
           \qquad \qquad $ - \eta\lambda_3 (\boldsymbol{L}_Z\times \boldsymbol{C})_{k*}$\\
           \qquad \qquad $ - \eta (\boldsymbol{\mathcal{B}}_{ijkl} - \boldsymbol{\mathcal{A}}_{ijkl}) \times \boldsymbol{\mathcal{G}} \times_1 \boldsymbol{S} \times_2 \boldsymbol{D} \times_4 \boldsymbol{T}$;
           
           \STATE $\boldsymbol{T}_{l*} = \boldsymbol{T}_{l*} - \eta\lambda_5 \boldsymbol{T}_{l*} - \eta\lambda_4 (\boldsymbol{T}_{l*} - \boldsymbol{T}_{l*} \times \boldsymbol{U}) \times (\boldsymbol{I} - \boldsymbol{U})^T$\\
           \qquad \qquad $ - \eta (\boldsymbol{\mathcal{B}}_{ijkl} - \boldsymbol{\mathcal{A}}_{ijkl}) \times \boldsymbol{\mathcal{G}} \times_1 \boldsymbol{S} \times_2 \boldsymbol{D} \times_3 \boldsymbol{C}$;
           
           \STATE $\boldsymbol{\mathcal{G}} = \boldsymbol{\mathcal{G}} - \eta\lambda_5 \boldsymbol{\mathcal{G}}$ \\
                 \qquad $- \eta (\boldsymbol{\mathcal{B}}_{ijkl} - \boldsymbol{\mathcal{A}}_{ijkl})\times \boldsymbol{S}_{i*} \circ \boldsymbol{D}_{j*} \circ \boldsymbol{C}_{k*} \circ \boldsymbol{T}_{l*};$
        \ENDFOR
    \STATE $loss_b = \mathcal{L}(\boldsymbol{\mathcal{G}}, \boldsymbol{S}, \boldsymbol{D}, \boldsymbol{C}, \boldsymbol{T})$;
    \UNTIL{$| loss_a - loss_b | < \epsilon $}
    \STATE $\boldsymbol{\widehat{\mathcal{A}}} = \boldsymbol{\mathcal{G}} \times_1 \boldsymbol{S} \times_2 \boldsymbol{D} \times_3 \boldsymbol{C} \times_4 \boldsymbol{T}$;
  \end{algorithmic}
\end{algorithm}

\subsection{Time Window based Parallel Decomposition Algorithm (TWPDA)}

Recent researches reveal that information diffusions in social networks exhibit a property of \textit{temporal locality}, i.e., a meme one focuses on at a time point is more similar to the memes she/he focuses on at close time points than those at distant time points \cite{r22,r23,r24}. For example, one may especially focus on a specific movie in May, and in June, and may still care about the comments about the movie, but she/he is unlikely to spread the memes about this movie in December. The temporal locality implies that in a CDT, a slice may has less to do with the slices at time points far from it than those at time points close to it. Inspired by this idea, we propose a Time Window based Parallel Decomposition Algorithm (TWPDA) for solving Equation (\ref{Eq_CIM}). 

TWPDA employs a sliding time window scheme, where the width of each time window is determined adaptively and separately. With respect to the time windows, TWPDA splits a CDT into a series of sub-tensors with different sizes along time dimension. Note that a CDT is a 4th-order tensor and its slice at time point $t_q$, $\boldsymbol{\mathcal{A}}_{***q}$, $1 \le q \le Q$, is a 3rd-order tensor, as we have defined in Section 3.1


\subsubsection{Sliding Time Window And Sub-tensor}
Let $T = \langle t_1, t_2, $ $\cdots, t_Q \rangle$ be a sequence of $Q$ time points considered, and then a sliding time window over $T$ and its corresponding sub-tensor can be defined as follows:
\begin{definition}
A \textit{sliding time window} is an interval, denoted by $[s, q]$, starting in time point $t_s$ and ending in time point $t_q$, $1 \le s \le q \le Q$, such that if $[s', q']$ is the successive time window of $[s, q]$, $s' = s+1$ and $q' > q$.
\label{Def_TW}
\end{definition}


\begin{figure}[tb]
\centering
    \epsfig{file=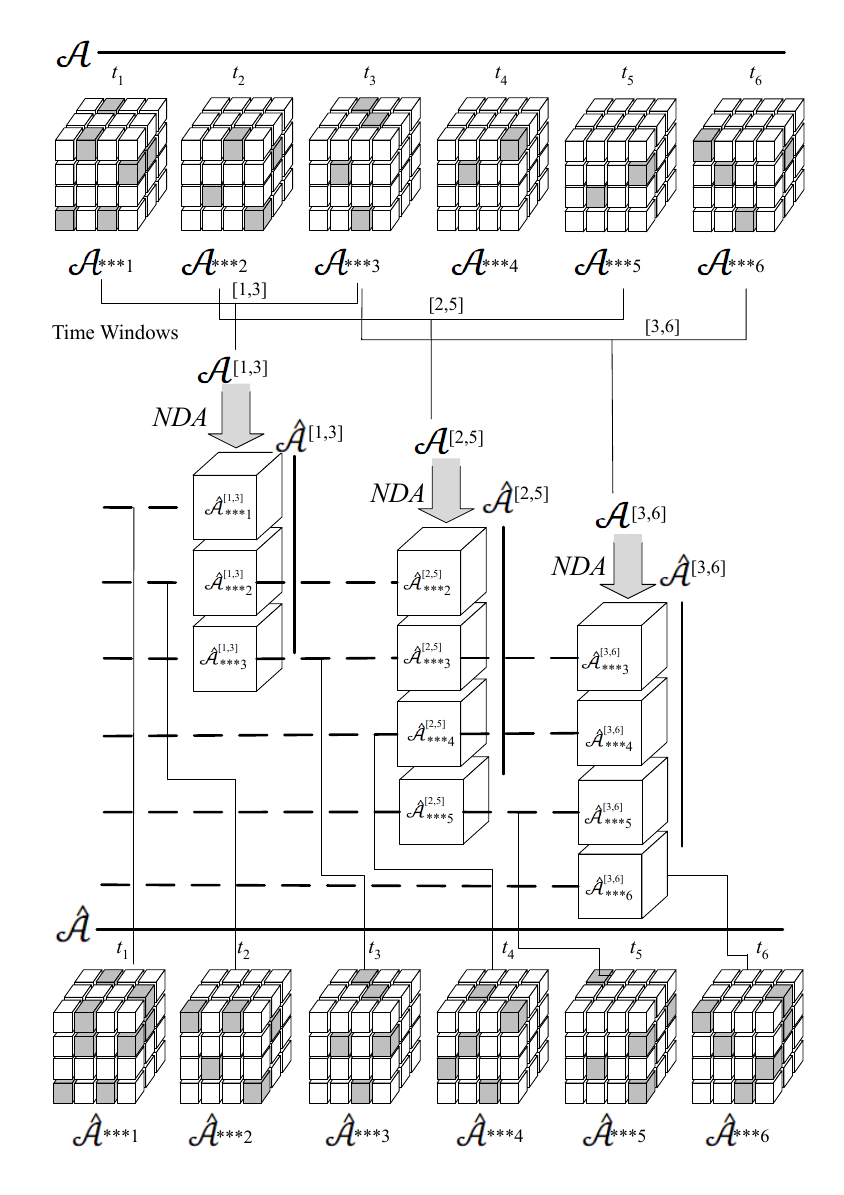}
\caption{Illustration of TWPDA.}
\label{Fig_TWPDA}
\end{figure}

\begin{definition}
The \textit{sub-tensor} corresponding to time window $[s,q]$ ($1 \le s \le q \le Q$), denoted by $\boldsymbol{\mathcal{A}}^{[s,q]}$, consists of the $q-s+1$ slices $\boldsymbol{\mathcal{A}}_{***i}$ ($i = s, s+1, \cdots, q$) concatenated along the time dimension.
\end{definition}

For example, in Fig. \ref{Fig_TWPDA}, there are 3 sliding time windows, $[1,3]$, $[2, 5]$, and $[3, 6]$, and their corresponding sub-tensors are $\boldsymbol{\mathcal{A}}^{[1,3]}$, $\boldsymbol{\mathcal{A}}^{[2,5]}$, and $\boldsymbol{\mathcal{A}}^{[3,6]}$. It is obvious that successive sub-tensors are overlapped, i.e., if $\boldsymbol{\mathcal{A}}^{[s',q']}$ is the successive sub-tensor of $\boldsymbol{\mathcal{A}}^{[s,q]}$, they share the slices from $\boldsymbol{\mathcal{A}}_{***s'}$ to $\boldsymbol{\mathcal{A}}_{***q}$. For example, in Fig. \ref{Fig_TWPDA}, $\boldsymbol{\mathcal{A}}^{[2,5]}$ is the successive sub-tensor of $\boldsymbol{\mathcal{A}}^{[1,3]}$, and they share the slices $\boldsymbol{\mathcal{A}}_{***2}$ and $\boldsymbol{\mathcal{A}}_{***3}$. Similarly, $\boldsymbol{\mathcal{A}}^{[3,6]}$ is the successive sub-tensor of $\boldsymbol{\mathcal{A}}^{[2,5]}$, and they share the slices from $\boldsymbol{\mathcal{A}}_{***3}$ to $\boldsymbol{\mathcal{A}}_{***5}$. 

\subsubsection{Determining Time Window Width}
As we have mentioned before, the observations of information diffusions are sparse. At the same time, due to the temporal locality of information diffusions, the distribution of the observations is likely nonuniform. Therefore, if the time windows share a fixed width, the number of available observations may vary drastically between different time windows, which results in that some sub-tensors may have insufficient data for inference while some ones may possess an excessive amount of data leading to overfitting. To address this issue, our idea is to adaptively determine the width of a time window with respect to the number of non-zero cells that the corresponding sub-tensor has, so that the denser (sparser) the observed infections are, the narrower (wider) the time window is. At the same time, we can note that the number of sub-tensors is also determined adaptively, since the sub-tensors are generated one by one with adaptively determined width.

Let $ nnz(\boldsymbol{\mathcal{A}}^{[s,q]}) $ be the number of non-zero cells of sub-tensor $\boldsymbol{\mathcal{A}}^{[s,q]}$. Then the width of the time window starting in time point $t_i$, denoted by $\omega_i$, is evaluated by the following function:
\begin{equation}
\omega_i = f(\alpha_i) =
\begin{cases}
Q - i + 1,  & \alpha_i \ge  Q - i + 1 \\
\alpha_i,  & nnz(\boldsymbol{\mathcal{A}}^{[i,(i+\alpha_i-1)]}) \ge \beta \text{ ,} \\
f(\alpha_i+1),  & nnz(\boldsymbol{\mathcal{A}}^{[i, (i+\alpha_i-1)]}) < \beta
\end{cases}
\label{Eq_TW_Width}
\end{equation}
where $\alpha_i  = \omega_{i-1}$ is the initial width of the time window starting in time point $t_i$, and $\beta$ is the given threshold of the number of non-zero cells. 

Equation (\ref{Eq_TW_Width}) shows that the time window starting at $t_i$ is produced by a growth from a width of at least $\alpha_i$. If $\alpha_i < Q+i-1$ and the corresponding sub-tensor has non-zero cells less than $\beta$, the time window grows via recursively invoking the function $f(\alpha_i)$ with $\alpha_i = \alpha_i+1$, until $\alpha_i \ge Q-i+1$ or the number of the non-zero cells of the corresponding sub-tensor (which is growing too) is greater than $\beta$. Whenever $\alpha_i \ge Q-i+1$, the width of that time window will be $Q-i+1$ regardless of the number of non-zero cells, and it will be the last time window. Note that the time window that ends at the last time point for the first time is the last time window, due to the following theorem:

\newtheorem{theorem}{Theorem}
\begin{theorem}
Given two successive sliding time windows $[s, q]$ and $[s+1, q']$, if $t_Q$ is the last time point and $q'=Q$, $q < Q$ and $[s+1, q']$ is the last time window. 
\label{The_TW_Width}
\end{theorem}
\begin{IEEEproof}
At first, according to Definition \ref{Def_TW}, $q \le Q$ and $q \ne Q$ since already $q'=Q$, so $q < Q$. Suppose there is one more time window $[s'+2,q'']$, and then $q'' = Q$ because $t_Q$ is the last time point. At the same time, $q'' \ne q' = Q$ again according to Definition \ref{Def_TW}, which is contradiction. Therefore there is no more time window and $[s+1, q']$ is exactly the last time window. 
\end{IEEEproof}

Fig. \ref{Fig_TWPDA} shows some time windows with different widths. In Fig. \ref{Fig_TWPDA}, the time window starting in $t_2$ and the time window starting in $t_3$ are stretched, since the slices $\boldsymbol{\mathcal{A}}_{***4}$ and $\boldsymbol{\mathcal{A}}_{***5}$ are pretty sparse. Note that $[3,6]$ is the last time window, since $[3,6]$ already includes the last time point $t_6$.

\subsubsection{Generating Final Result}
As we have mentioned, due to the sliding of time windows, a time slice at a specific time point may be often shared in two or more sub-tensors, which results in that more than one sub-tensors will separately produce the approximations of the shared slice. For example, in Fig. \ref{Fig_TWPDA}, the slice $\boldsymbol{\mathcal{A}}_{***3}$ simultaneously appears in the sub-tensors $\boldsymbol{\mathcal{A}}^{[1,3]}$, $\boldsymbol{\mathcal{A}}^{[2,5]}$, and $\boldsymbol{\mathcal{A}}^{[3,6]}$, and the approximations of these sub-tensors, $\boldsymbol{\widehat{\mathcal{A}}}^{[1,3]}$, $\boldsymbol{\widehat{\mathcal{A}}}^{[2,5]}$, and $\boldsymbol{\widehat{\mathcal{A}}}^{[3,6]}$, have their respective $\boldsymbol{\widehat{\mathcal{A}}}_{***3}$, the approximation of $\boldsymbol{\mathcal{A}}_{***3}$. 

Here we need to reasonably generate the final result for a shared slice from its multiple approximations. Our idea to address this issue is taking the weighted average of the multiple approximations of a shared slice as its final result that will appear in the final approximated CDT, and according to the temporal locality, it is reasonable that the wider the time window, the smaller the weight of the approximation produced by the sub-tensor corresponding to that time window. Based on this idea, the final approximation of a time slice $\boldsymbol{\mathcal{A}}_{***k}$ is evaluated as 
\begin{equation}
\boldsymbol{\widehat{\mathcal{A}}}_{***k} = \frac {1}{W}\sum_{\boldsymbol{\widehat{\mathcal{A}}}^{[s,q]} \text{ s.t. } s \le k \le q} w^{[s,q]} \times \boldsymbol{\widehat{\mathcal{A}}}^{[s,q]}_{***k},
\label{Eq_Final_Result}
\end{equation}
where $\boldsymbol{\widehat{\mathcal{A}}}^{[s,q]}_{***k}$ is the approximation of $\boldsymbol{\mathcal{A}}_{***k}$ that is contained in the approximated sub-tensor $\boldsymbol{\widehat{\mathcal{A}}}^{[s,q]}$, $w^{[s,q]} = \frac {1} {2^{|q-s+1|}}$ is the weight of $\boldsymbol{\widehat{\mathcal{A}}}^{[s,q]}_{***k}$, and $W = \sum w^{[s,q]}$ is the regularization factor.

\begin{algorithm}[t]
    \caption{ \emph{TWPDA}$(\boldsymbol{\mathcal{A}},\boldsymbol{X},\boldsymbol{Y},\boldsymbol{Z},\boldsymbol{U},R,\epsilon,\alpha_1,\beta)$ }
    \label{Alg_TWPDA}
    \begin{algorithmic}[1]
        \REQUIRE ~~
           $\boldsymbol{\mathcal{A}}$: the sparse CDT; 
           $\boldsymbol{X}$: the SDA matrix; 
       	    $\boldsymbol{Y}$: the NMA matrix; 
           $\boldsymbol{Z}$: the MC matrix; 
           $\boldsymbol{U}$: the TS matrix; 
           $R$: the target rank; 
           $\epsilon$: the threshold of error; 
           $\alpha_1$: the initial width of first time window; 
           $\beta$: the threshold of the number of non-zero cells;
        \ENSURE ~~ 
           $\boldsymbol{\widehat{\mathcal{A}}}$: the approximated CDT;
           
        \STATE Initialize the cells of $\boldsymbol{\widehat {\mathcal{A}}}$ with zero;
        \STATE Generate the sliding time windows $\{ [s,q] \}$ by using Equation (\ref{Eq_TW_Width}) with $\alpha_1$ and $\beta$;
        \STATE Construct the sub-tensors $\{ \boldsymbol{\mathcal{A}}^{[s,q]} \}$;
        \STATE Generate $\{ \boldsymbol{\widehat{\mathcal{A}}}^{[s,q]}\}$ in parallel, by invoking $NDA$ with $\boldsymbol{\mathcal{A}}^{[s,q]}$, $\boldsymbol{X}$, $\boldsymbol{Y}$, $\boldsymbol{Z}$, $\boldsymbol{U}^{[s,q]}$, $R$ and $\epsilon$, where $\boldsymbol{U}^{[s,q]}$ is the sub-matrix consisting of rows from $\boldsymbol{U}_{s*}$ to $\boldsymbol{U}_{q*}$;
        \STATE Generate in parallel the final result $\boldsymbol{\widehat{\mathcal{A}}}_{***k}$, $1 \le k \le Q$, according to Equation (\ref{Eq_Final_Result});
        \STATE Generate $\boldsymbol{\widehat{\mathcal{A}}}$ by concatenate the approximated time slices$\{ \boldsymbol{\widehat{\mathcal{A}}}_{***k} \}$ along the time dimension;   \end{algorithmic}
\end{algorithm}

\subsubsection{Outline of TWPDA}
The outline of TWPDA is given in Algorithm \ref{Alg_TWPDA}, where the major computation is invoking NDA (shown in Algorithm \ref{Alg_NDA}) in parallel for each sub-tensor to generate the  approximation (Line 4). Note that for a sub-tensor $ \boldsymbol{\mathcal{A}}^{[s,q]} $, the temporal smoothing matrix defined in Equation (\ref{Eq_TS}) becomes a sub-matrix $\boldsymbol{U}^{[s,q]}$ of $\boldsymbol{U}$, since $[s,q]$, is just a subinterval of $[1,Q]$. Compared with NDA, the computational complexity of an iteration of the sub-tensor $ \boldsymbol{\mathcal{A}}^{[s,q]} $ drops to $O(\frac{Q}{q-s+1}n)$, because the number of nonzero cells has been reduced to $\frac{Q}{q-s+1}n$, where $n$ denotes the number of nonzero elements of the original CDT.

\begin{table}[tb]
\centering
\caption{Synthetic Datasets}
\label{TBL_SYNDATA}
\begin{tabular}{cccc} 
\hline
Name & $N$ & $M$ & $Q$ \\ 
\hline
SYN1 & 1K & 2 & 35 \\ 
\hline
SYN2 & 2K & 2 & 35 \\ 
\hline
SYN3 & 3K & 2 & 35 \\ 
\hline
SYN4 & 4K & 2 & 35 \\
\hline
SYN5 & 5K & 2 & 35 \\
\hline
\end{tabular}
\end{table}

\section{Experiments}
In this section, we present the experimental results on real world datasets and synthetic datasets. The experiments are conducted on a Spark cluster consisting of 3 PCs where each PC equipped with a 2.7 GHz INTEL CPU of 4 cores and 128GB RAM, and all the programs are written with MATLAB 2016b. 

\subsection {Experiment Setting}

\subsubsection{Datasets}

\textbf{Twitter} We choose the Twitter dataset released by Zhang \textit{et al.} \cite{r29}, which consists of 8 million tweets posted by 5,000 users during the period from Jan. 2009 to Dec. 2012. From the Twitter dataset, we extract 5 memes and construct a CDT of $\mathbb{R}^{5000 \times 5000 \times 5 \times 48}$ for test, where the time dimension consists of 48 months. We use the Foursquare dataset also released by Zhang \textit{et al.} \cite{r29} as the additional data source, which shares the users with the Twitter dataset and from which we build the heterogeneous constraint matrices $\boldsymbol{X}$, $\boldsymbol{Y}$ and $\boldsymbol{Z}$.

\textbf{Weibo} The Sina Weibo network is the largest microblogging website in China. The first Weibo dataset is crawled with the method proposed Kong \textit{et al.} \cite{r30}, which consists of 80,000 tweets posted by 3,000 users in May, 2014. From the first Sina Weibo dataset, we also extract 5 memes and construct a CDT of $\mathbb{R}^{3000 \times 3000 \times 5 \times 31}$ for test, where the time dimension consists of 31 days. We also crawled the second Sina Weibo dataset for the same users as the additional data source from which we build the heterogeneous constraint matrices $\boldsymbol{X}$, $\boldsymbol{Y}$ and $\boldsymbol{Z}$. In the second Weibo dataset, the tweets are posted in April 2014, which are before the time of the tweets in the first Sina Weibo dataset.

\textbf{Synthetic Datasets} In order to investigate the scalability of TWPDA, we generate 5 synthetic datasets as described in Table \ref{TBL_SYNDATA}. For each synthetic dataset, we build a synthetic CDT of $ \mathbb{R}^{N\times N\times M \times Q} $, where $N$ is the number of network nodes, $M$ is the number of memes diffusing over the network, and $Q$ is the number of time points. The value of a synthetic CDT cell is random generated according to the standard normal distribution.

\subsubsection{Metrics}

To evaluate the effectiveness of the proposed CIM, we use two metrics, recovery accuracy (RA) and root mean square error (RMSE). Let $\boldsymbol{\mathcal{A}}_{ijmq}$ be a non-zero cell whose original value is removed as the ground truth and $\boldsymbol{\widehat{\mathcal{A}}}_{ijmq}$ be its estimate, and then RA and RMSE are defined as follows:
\begin{equation}
RA = \frac{1}{S}\sum_{i,j,m,q}\mathbb{I}(\boldsymbol{\mathcal{A}}_{ijmq}\boldsymbol{\widehat{\mathcal{A}}}_{ijmq}),
\label{Eq_RA}
\end{equation}
\begin{equation}
RMSE = \sqrt{\frac{1}{S}\sum_{i,j,m,q}(\boldsymbol{\mathcal{A}}_{ijmq}-\boldsymbol{\widehat{\mathcal{A}}}_{ijmq})^2},
\label{Eq_RMSE}
\end{equation}
where $S$ is the number of cells removed for test, and $\mathbb{I}(x) = 1$ if $x > 0$, otherwise $\mathbb{I}(x) = 0$.

Basically, RA measures the ability to qualitatively detect an unobserved infection which does exist, while RMSE evaluates the accuracy of the estimation of how many times that infection occurs. 

\subsubsection{Baseline Methods} 
We use two existing representative algorithms, NETINF \cite{r27} and SSR \cite{r48}, as the baseline methods. NETINF reconstructs an underlying network based on observations from multiple cascades using a greedy strategy, while SSR infers a diffusion history based on maximum likelihood estimation. We apply them to reconstruct the multiple testing diffusion traces one by one, and evaluate their respective performance by the average of the results over the testing diffusion traces. 

To verify the effectiveness of the heterogeneous constraints ($\boldsymbol{X}$, $\boldsymbol{Y}$, $\boldsymbol{Z}$ and $\boldsymbol{T}$), we also compare CIM to Tucker Decomposition (TD) \cite{r06} and the combinations of TD and subsets of the constraints, i.e., TD+X, TD+X+Y, and TD+X+Y+Z, where each method adds one more constraint to the previous one. TD is a classical algorithm for tensor decomposition, which decomposes a \textit{n}th-order tensor into a product of a core tensor and $n$ factor matrices. 



\subsection{Effectiveness of CIM}

\begin{figure}[tb]
\centering
    \begin{minipage}[lb]{0.20\textwidth}
        \centering
         \subfigure[ ]{\includegraphics[width=3.7cm,height=3.2cm]{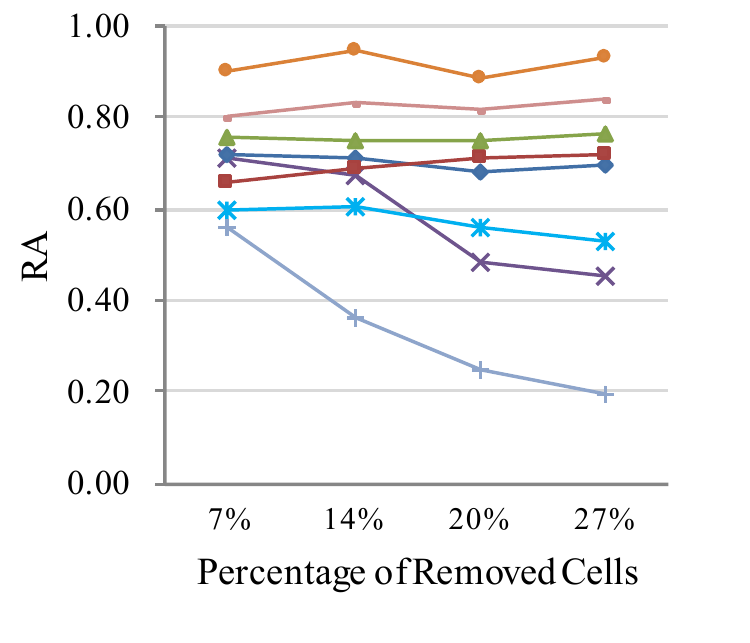}}
    \end{minipage}
    \begin{minipage}[lb]{0.25\textwidth}
         \centering
         \subfigure[ ]{\includegraphics[width=4.7cm,height=3.2cm]{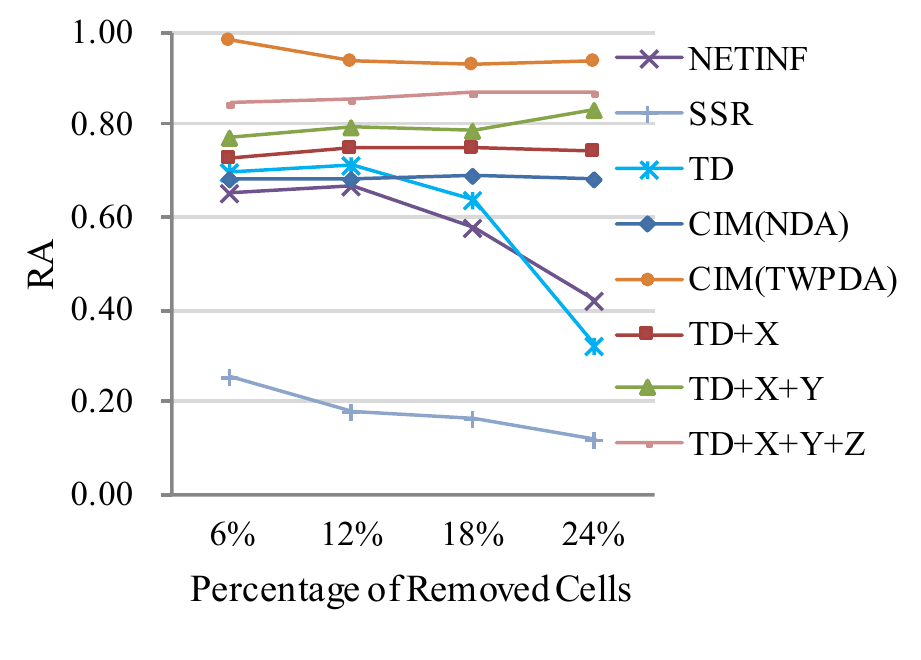}}
    \end{minipage}
\caption{RA on (a) Twitter and (b) Weibo}
\label{Fig_Accuracy}
\end{figure}

\begin{figure}[tb]
\centering
    \begin{minipage}[lb]{0.21\textwidth}
        \centering
         \subfigure[ ]{\includegraphics[width=3.9cm,height=3.1cm]{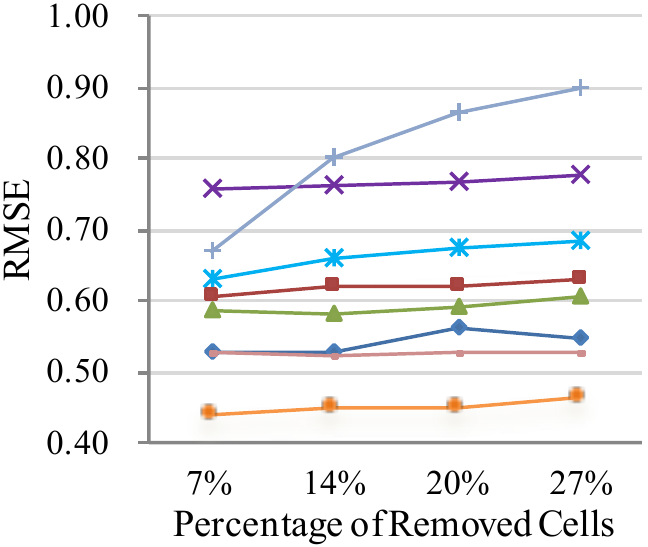}}
    \end{minipage}
    \begin{minipage}[lb]{0.25\textwidth}
         \centering
         \subfigure[ ]{\includegraphics[width=4.9cm,height=3.1cm]{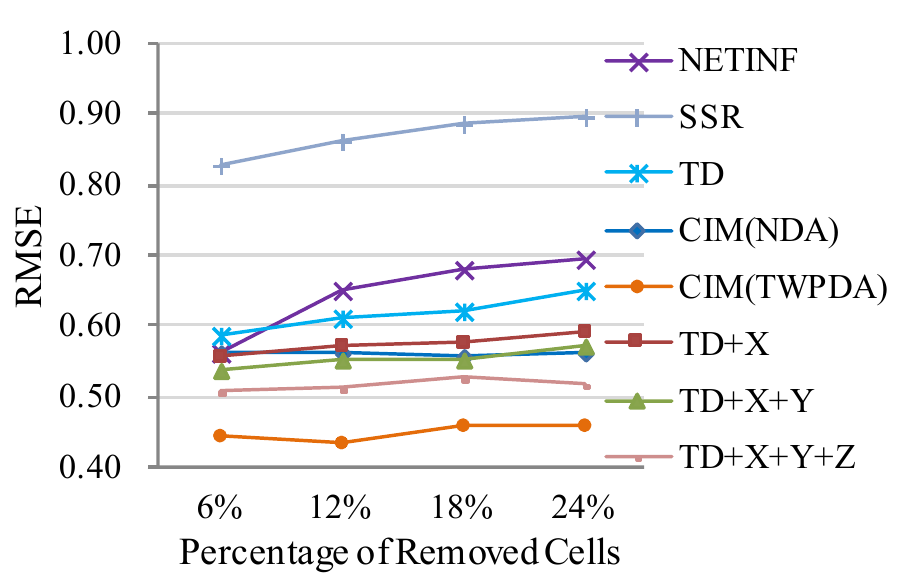}}
    \end{minipage}
\caption{RMSE on (a) Twitter and (b) Weibo}
\label{Fig_RMSE}
\end{figure}

To verify the effectiveness of the proposed CIM (with TWPDA and with NDA respectively), on the CDT of each dataset, we randomly remove different percentages of cells and use their original values as the ground truth. The setting of parameters is $R = 3$, $\epsilon = 0.01$, $\lambda_1 =  \lambda_2 = 1$, $\lambda_3 = 0.3$, $\lambda_4 = \lambda_5 = 0.05$, and $\eta = 0.001$. Fig. \ref{Fig_Accuracy} and Fig. \ref{Fig_RMSE} show the RA and RMSE of CIM and the baseline methods, respectively, from which we have the following observations:
\begin{compactenum}[(1)]

\item On each scale of the removed cells of both Twitter and Weibo, the RA and RMSE of CIM (with TWPDA or NDA) are obviously higher and lower than those of the baseline methods, respectively, which indicates that CIM can recover unobserved infections more correctly and also estimate their frequency more accurately. This result can be explained by two reasons. First, CIM attaches the importance to the coexisting information diffusions and collaboratively infers them via a low-rank approximation of a CDT, which takes advantage of the synergism between the coexisting information diffusions. Second, by fusing the additional heterogeneous constraints into the decomposition of CDT, CIM reduces the uncertainty of CDT significantly.

\item We can also note that with the increase of the scale of the removed cells, the performance of CIM keeps approximately stable, while the performance of alternative methods drops markedly. Since the increasing number of the removed cells leads to the sparser data, this result indicates that CIM outperforms the alternative methods on sparse data due to its collaborative reconstruction which utilizes the correlation of coexisting information diffusions via the low-rank approximation of sparse CDT. 

\end{compactenum}

\begin{figure}[tb]
\centering
    \centering
    \begin{minipage}[lb]{0.20\textwidth}
        \centering
         \subfigure[ ]{\includegraphics[width=3.5cm,height=3.2cm]{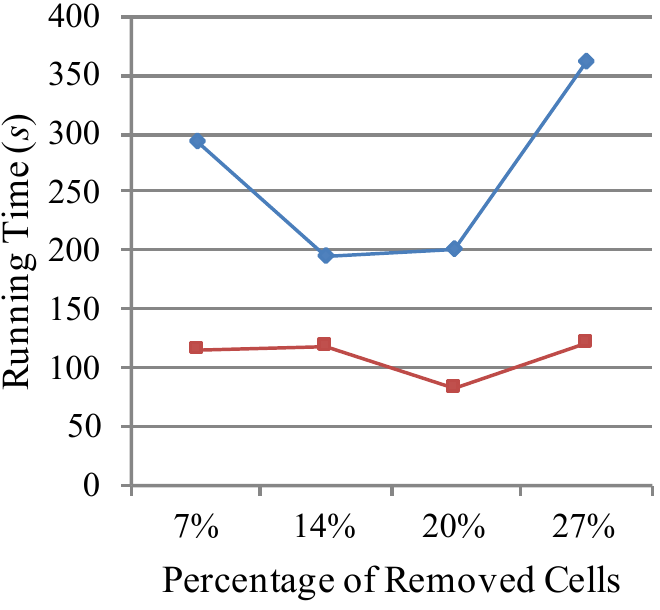}}
    \end{minipage}
    \begin{minipage}[lb]{0.25\textwidth}
         \centering
         \subfigure[ ]{\includegraphics[width=3.5cm,height=3.2cm]{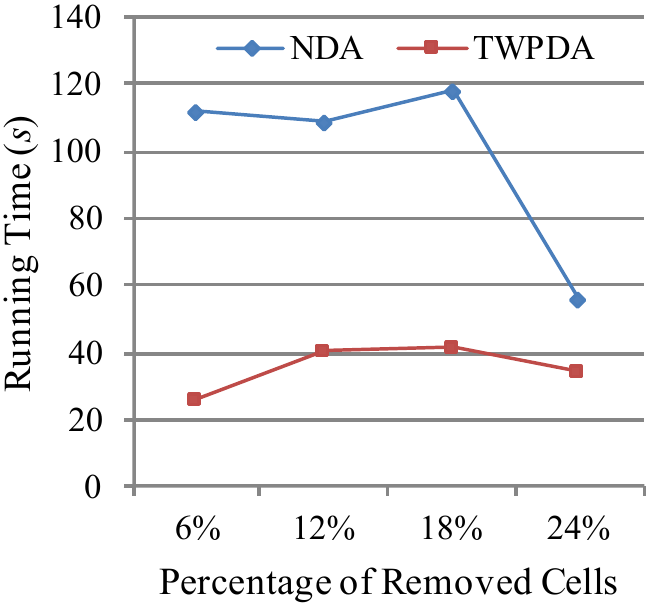}}
    \end{minipage}
\caption{Time Cost of NDA and TWPDA on (a) Twitter and (b) Weibo}
\label{Fig_TIME}
\end{figure}

\subsection{Effectiveness of Heterogeneous Constraints}

In Fig. \ref{Fig_Accuracy} and Fig. \ref{Fig_RMSE}, the curves of TD, TD+X, TD+X+Y, TD+X+Y+Z, and CIM show that the more number of constraints is taken into consideration, the higher the RA and the lower the RMSE. For example, on both datasets, TD+X+Y+Z is more accurate than TD+X+Y, while TD+X +Y is more accurate than TD+X and TD. As CIM takes into consideration all the 4 constraints, it outperforms all the others. These results show that fusing the heterogeneous constraints with the CDT decomposition does improve the accuracy of the inference of coexisting information diffusions. 

\subsection{Efficiency of TWPDA}

\subsubsection{TWPDA vs NDA}
At first, we can see from Fig. \ref{Fig_Accuracy} and Fig. \ref{Fig_RMSE} that on both Twitter and Weibo, the RA of CIM with TWPDA is higher than the RA of CIM with NDA, and the RMSE of CIM with TWPDA is lower than the RMSE of CIM with NDA, which confirms the effectiveness of the proposed sliding time window scheme taking into consideration the temporal locality of information diffusions. As we have described in Section 4, TWPDA splits a CDT into sub-tensors sharing slices with respect to a series of overlapped sliding time windows, and makes the final estimation of a shared slice in terms of the weighted mean of its value in each sub-tensor sharing it, which leads to a more accurate result than NDA as NDA performs the decomposition on the whole CDT without the consideration of the temporal locality of information diffusions. 

On the other hand, we can also unsurprisingly observe from Fig. \ref{Fig_TIME} that the running time of TWPDA is significantly less than that of NDA on both Twitter and Weibo. This efficiency gain of TWPDA is again due to the sliding time window scheme which enables a parallel decomposition of the sub-tensors. Note that the parallel degree of TWPDA depends on the number of sub-tensors, which is adaptively decided according to the density distribution of data, as described in Subsection 4.2.

In conclusion, the results shown in Fig. \ref{Fig_Accuracy}, Fig. \ref{Fig_RMSE}, and Fig. \ref{Fig_TIME} together confirm that TWPDA can speed up the decomopistion of CDT without compromise on accuracy due to its sliding time window scheme which is based on the temporal locality of information diffusions and leads to a parallel decomposition of CDT.

\begin{figure}[tb]
\centering
    \begin{minipage}[lb]{0.21\textwidth}
        \centering
         \subfigure[ ]{\includegraphics[width=3.6cm,height=3.2cm]{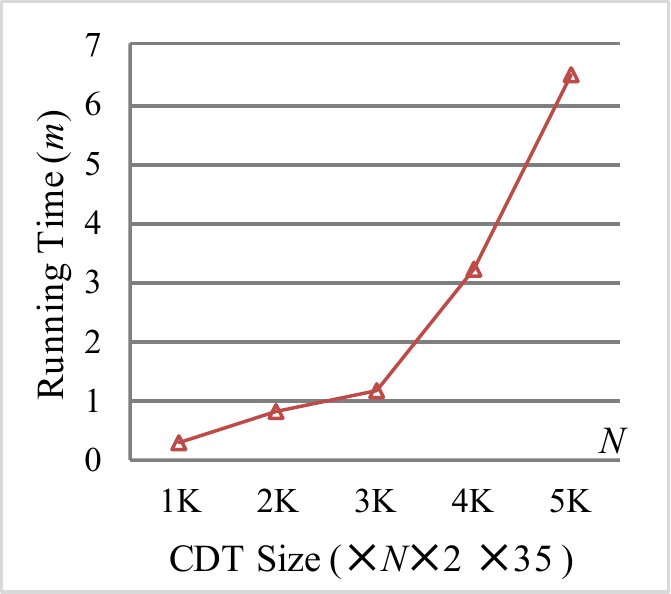}}
    \end{minipage}
    \begin{minipage}[lb]{0.25\textwidth}
         \centering
         \subfigure[ ]{\includegraphics[width=3.6cm,height=3.2cm]{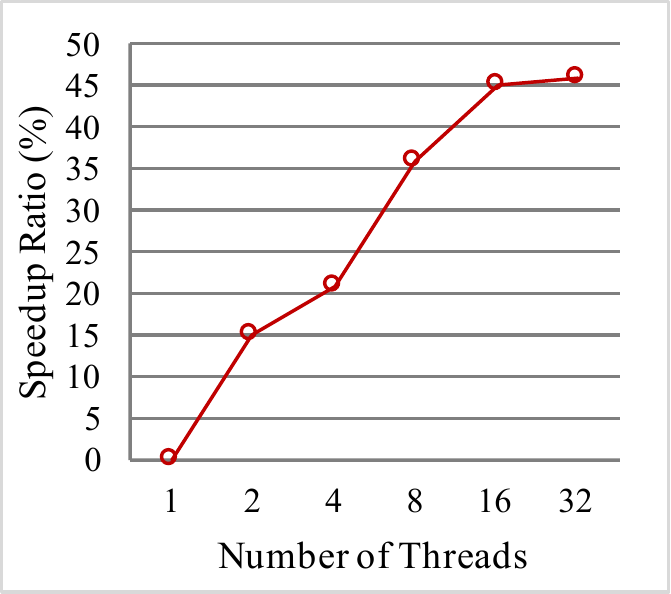}}
    \end{minipage}
\caption{Scalability of TWPDA: (a) Running time vs Size of CDT (b) Speedup Ratio vs Number of Sub-tensors}
\label{Fig_Scalability}
\end{figure}

\begin{figure}[tb]
\centering
    \centering
    \begin{minipage}[lb]{0.20\textwidth}
        \centering
         \subfigure[ ]{\includegraphics[width=3.8cm,height=3.2cm]{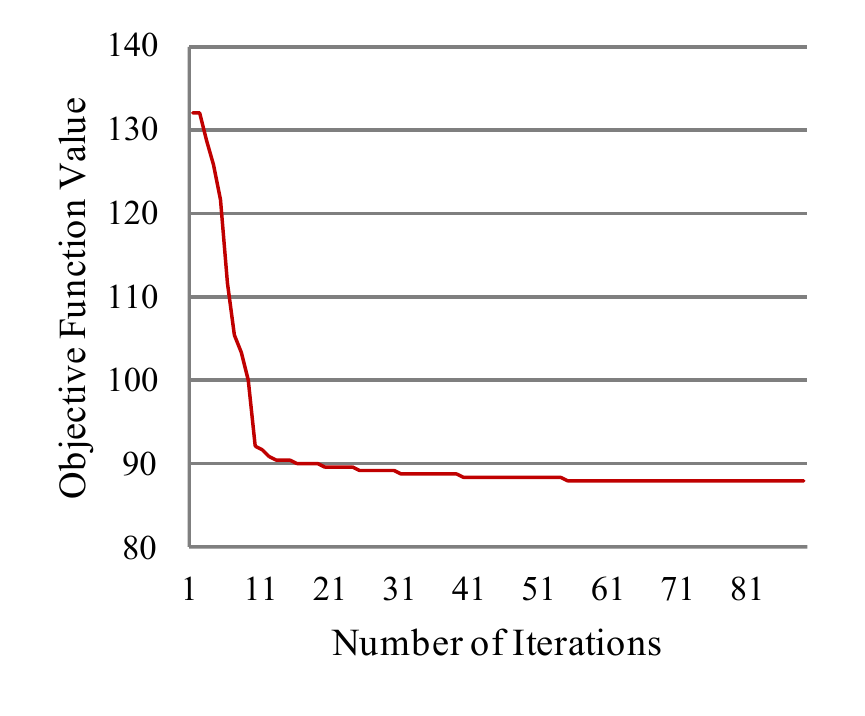}}
    \end{minipage}
    \begin{minipage}[lb]{0.25\textwidth}
         \centering
         \subfigure[ ]{\includegraphics[width=3.8cm,height=3.2cm]{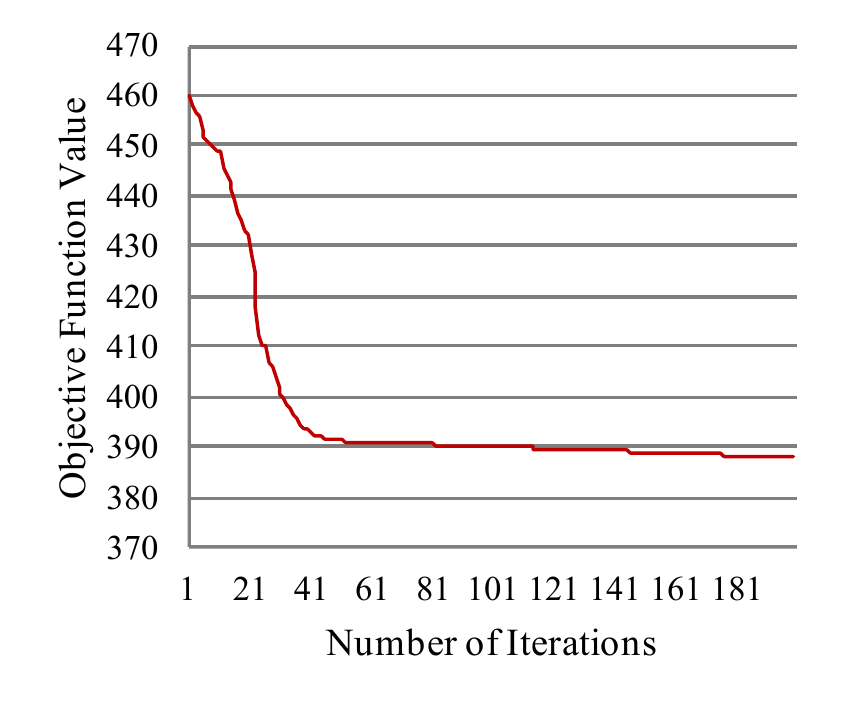}}
    \end{minipage}
\caption{Convergence of CIM on (a) Twitter and (b) Weibo}
\label{Fig_Convergence}
\end{figure}

\subsubsection{Parallelizability of TWPDA}

At first, we investigate the running time of TWPDA over the 5 synthetic datasets of different scales listed in Table \ref{TBL_SYNDATA}, where each synthetic CDT is divided into 12 overlapped sub-tensors, each of which is served by a thread running on a physically separate core. Fig. \ref{Fig_Scalability}(a) shows that as the size of CDT increases, the running time of TWPDA almost quadratically grows.

We also investigate the speedup ratio of TWPDA over different numbers of threads, each of which serves one sub-tensor, using the largest synthetic dataset SYN5. Fig. \ref{Fig_Scalability}(b) shows that the speedup ratio sharply rises from $0\%$ to about $45\%$ as the number of threads increases from 1 from 16, which means the speed of TWPDA accelerates as more threads are generated for parallel decomposition. When the number of threads is greater than 16, the speedup ratio converges, which indicates that the speedup gain is limited by the number of the physical CPU cores (in our platform which is 12).  


\subsection{Convergence of CIM}
As TWPDA invokes NDA for decompositions of sub-tensors, the convergence of CIM ultimately depends on the convergence of NDA. Fig. \ref{Fig_Convergence} shows the convergence curves of CIM with NDA. As we can see from Fig. \ref{Fig_Convergence}, the objective function (defined by Equation (\ref{Eq_Objective})) quickly converges to the vicinity of its minimum after about 10 iterations on Twitter and 40 iterations on Weibo.

\section{Related Work}
In this section, we briefly review the previous works related to diffusion inference that can be roughly categorized into two classes, i.e., network inference and diffusion history reconstruction.


\subsection{Network Inference}

In network inference, one might want to recover the underlying diffusion network structure by only observing the information cascades (i.e., the "infection times" of nodes) over it. There have been a variety of approaches proposed for network inference \cite{r27,r51,r53,r05,r54}. Gomez-Rodriguez \textit{et al} \cite{r27}. propose an approach called NETINF, which can be viewed as the groundbreaking and representative work for inferring the network connectivity. NETINF models an information cascade as a tree, and builds a generative model to recover the underlying diffusion network structure with a greedy strategy. Myers \textit{et al}. \cite{r51} present a maximum likelihood approach, called CONNIE, which infers the diffusion network based on convex programming with a $l_1$-like penalty term that encourages sparsity. Recently, some studies concern the network inference from sparse recovery perspective. Wang \textit{et al}. \cite{r05} propose an optimization framework called TrNetInf, which can transfer the structure knowledge from an external diffusion network with sufficient cascade data to help infer the hidden diffusion network with sparse cascades. Pouget-Abadie \textit{et al}. \cite{r53} introduce a generalized linear cascade model which formulates the network inference problem in the context of discrete-time influence cascades as a sparse recovery problem. Different from the above works which depend on a parametric model, Rong \textit{et al}. \cite{r54} propose an algorithm called Non-Parametric Distributional Clustering (NPDC). NPDC interprets the diffusion process from the cascade data directly in a non-parametric way, and infers the diffusion network according to the statistical difference of the infection time intervals between nodes connected with diffusion edges versus those with no diffusion edges.

\subsection{Diffusion History Inference}
Diffusion history inference is the most related to our work, which is fundamentally different from network inference. The methods for network inference essentially is to infer the unknown network edges (e.g., the following relationship between users of social media) often assuming full histories of diffusion traces are observable, while the purpose of diffusion history inference is to reconstruct the missing history (consisting of infections, e.g., commenting or retweeting a tweet) of a single diffusion trace.

Recently, some works on diffusion history inference have been reported. Sefer \textit{et al}. \cite{r47} reduce the problem of diffusion history inference to the problem of determining the maximum likelihood history given diffusion snapshots, with independent cascade (IC) assumption \cite{r14}, and propose an algorithm called DHR-sub (sub-modular history reconstruction on discrete dynamics) which reconstructs the history by greedily maximizing the non-monotone sub-modular log-likelihood at each time step. Chen \textit{et al}. \cite{r48} formulate the diffusion history inference problem as a maximum a posteriori (MAP) estimate problem, and propose a greedy and step-by-step reconstruction algorithm called SSR to infer the most likely historical diffusion trace. Rozenshtein \textit{et al}. \cite{r25} consider this problem in a different way, which models an information diffusion as a temporal Steiner-tree, and recover its historical spread flow by searching a temporal Steiner-tree with minimum cost defined in advance. 


The biggest difference between the Collaborative Inference Model (CIM) proposed in this paper and the existing methods lies in that by exploiting the phenomena that in real world multiple diffusion traces of different information often coexist over the same social network, CIM can model multiple diffusion traces holistically and recover their missing histories in a synergistic way. 

\section{Conclusion}

In this paper, we propose a novel approach called CIM for the problem of the inference of coexisting information diffusions histories. CIM can holistically model multiple information diffusion traces as a sparse 4th-order tensor called Coexisting Diffusions Tensor (CDT), and can collaboratively infer the histories of the coexisting information diffusions via a low-rank approximation of CDT with a fusion of heterogeneous constraints generated from additional data sources. To speed up the inference without compromise on the accuracy, we further propose an optimal algorithm called Time Window based Parallel Decomposition Algorithm (TWPDA) as the core component of CIM. TWPDA divides a CDT into sub-tensors along the time dimension by utilizing the temporal locality of information diffusions, and fulfills the decomposition in parallel. The extensive experiments conducted on real world datasets and synthetic datasets verify the effectiveness and scalability of CIM and TWPDA.



\bibliographystyle{IEEEtran}
\bibliography{IEEEabrv,CIM}

\end{document}